\documentclass[showpacs,nofootinbib,preprintnumbers,amsmath,amssymb,preprint,aps]{revtex4}
\usepackage[dvips]{graphicx}
\usepackage{graphicx}
\usepackage{amsfonts}
\usepackage{bm}
\usepackage{amsmath}
\usepackage{amssymb}
\usepackage{color}
\usepackage[all]{xy}

\def\be{\begin{equation}}
\def\ee{\end{equation}}
\def\bea{\begin{eqnarray}}
\def\eea{\end{eqnarray}}

\begin{document}

\title{Multipole structure of compact objects}
\author{Hernando Quevedo}
\email{quevedo@nucleares.unam.mx}
\affiliation{Instituto de Ciencias Nucleares, Universidad Nacional Aut\'onoma de M\'exico, AP 70543, M\'exico, DF 04510, Mexico \\
Dipartimento di Fisica and ICRA, Universit\`a di Roma ``La Sapienza'', I-00185 Roma, Italy}



\begin{abstract}
We analyze the applications of general relativity  in relativistic astrophysics in order to solve the problem of describing 
the geometric and physical properties of the interior and exterior gravitational and electromagnetic fields of compact objects. We focus on the 
interpretation of exact solutions of Einstein's equations in terms of their multipole moments structure. In view of the lack of physical interior
solutions, we propose an alternative approach in which higher multipoles should be taken into account. 


\end{abstract}


\maketitle

\section{Introduction}

Einstein's equations \cite{eins15}
\be
R_{\mu\nu} - \frac{1}{2} R g_{\mu\nu} = \kappa T_{\mu\nu} 
\label{eins}
\ee
relate the geometric structure of spacetime, which is given by the Einstein tensor $G_{\mu\nu}= R_{\mu\nu} - \frac{1}{2} R g_{\mu\nu}$, with the 
matter content of spacetime, which is given by the energy-momentum tensor  $ T_{\mu\nu} $. Moreover, the equivalence principle allows us to establish
a relationship between the geometric structure of spacetime and the gravitational field. This implies that the curvature of spacetime, interpreted as four-dimensional 
differential Riemannian manifold, can be considered as a measure of the gravitational interaction. For instance, if the spacetime is flat, $R_{\mu\nu\lambda\tau}=0$, 
the Einstein tensor vanishes, leading to the condition that the matter content of the spacetime vanishes too. The opposite is not true. If $T_{\mu\nu}=0$, the Ricci 
tensor $R_{\mu\nu}$ and the curvature scalar $R$ vanish as well, but the curvature tensor  $R_{\mu\nu\lambda\tau}$ can be different from zero.

Einstein's general relativity is thus a theory of the gravitational interaction. As such, it should be able to describe all physical situations in which the gravitational field is involved. Consider, for instance, the case of astrophysical compact objects, i.e., objects that are small for their mass. In general, 
the class of astrophysical compact objects is often considered to contain collectively planet-like objects, white dwarfs, neutron stars, other exotic dense stars, 
and black holes. The problem of describing the gravitational field of compact objects can be split into two related problems, namely, 
the exterior and the interior field, each of them represented by a particular metric $g_{\mu\nu}^i $ and $g_{\mu\nu}^e$, respectively. 
The surface of the compact object represents the hypersurface at which the interior and the exterior fields must coincide. The exterior field corresponds to a
vacuum spacetime $T_{\mu\nu}=0$, for which Einstein's equations reduce to 
\be
R_{\mu\nu} =0 \ ,
\ee
whereas for the interior field it is necessary to choose a particular energy-momentum tensor that would take into account 
all the physical properties of the internal structure of the compact object. A particularly simple choice is the perfect-fluid energy-momentum tensor \cite{solutions}
\be
T_ {\mu\nu} = (\rho + p )u_\mu u_\nu - p g_{\mu\nu}  \ ,
\label{pf}
\ee
where $\rho$ and $p$ are the density and the pressure of the fluid, respectively, and $u^\mu$ is the 4-velocity. The case of fluids with anisotropic pressures is also used in the literature. 

Most compact objects, however, are characterized by the presence of internal and external electromagnetic fields. This implies that in general we should consider the Einstein-Maxwell theory. In this case, the exterior field should be described by an exterior metric which satisfies the electrovacuum equations
\be
R_{\mu\nu} = \kappa\left( F_{\lambda\mu} F^{\lambda}_{\ \nu} - \frac{1}{4} g_{\mu\nu} F_{\lambda\tau}F^{\lambda\tau}\right) \ , \quad 
F^{\mu\nu}_{\ \ ;\mu} = 0 \ , 
\label{emax1}
\ee
where $F_{\mu\nu}$ is the Faraday electromagnetic tensor. On the surface of the compact object, the exterior metric must be matched with the interior metric 
which satisfies the general equations
\be
R_{\mu\nu} - \frac{1}{2} R g_{\mu\nu} = \kappa  [(\rho + p )u_\mu u_\nu - p g_{\mu\nu} ]
+ \kappa \left( F_{\lambda\mu} F^{\lambda}_{\ \nu} - \frac{1}{4} g_{\mu\nu} F_{\lambda\tau}F^{\lambda\tau}\right) \ , 
\ee
\be 
F^{\mu\nu}_{\ \ ;\mu}  = 0 \ .
\ee

In this work, we present a brief review of the main exact solutions of Einstein-Maxwell equations which can be used to describe the exterior and interior field of astrophysical compact objects. We interpret the solutions in terms of their multipole solutions. To this end, we use the Geroch-Hansen \cite{ger70,hans74}
procedure which provides a relativistic and coordinate-invariant definition of multipole moments. According to this definition, any stationary, vacuum, 
asymptotically flat solution of Eistein's equations can be uniquely characterized by two sets of multipoles $M_n$ and  $J_n$ ($n=0,1,..)$, the first of which 
represents the field generated by the mass distribution whereas the second one is due to  the rotation of the distribution. In the case of electrovacuum fields, 
two more sets  $E_n$ and $H_n$ must be included, representing the electric and magnetic multipoles of the electromagnetic field, respectively. The concrete 
calculations necessary to find the explicit expressions of the multipole moments for particular solutions are not easy to be carried out. Some auxiliary 
procedures can be used which are based on particular representations of the solutions. Here, we will use the Ernst representation and the concrete formulas derived in \cite{quev90,quev92}.

This paper is organized as follows. First, in Sec. \ref{sec:lel}, we present the general line element that can be used to investigate the structure of the exterior and interior field equations. In Sec. \ref{sec:ext}, we present the Kerr metric with its multipole structure and its  generalizations which include the electric charge and
higher multipole moments. In Sec. \ref{sec:int}, we mention the main problems associated with the search for perfect-fluid interior solutions of Einstein and Einstein-Maxwell equations. We propose an alternative approach that implies the inclusion of the quadrupole as an additional degree of freedom. This could contribute to solve 
the general set of Einstein-Maxwell equations. Section \ref{sec:con} contains the conclusions.   We use throughout the paper geometric units with $G=c=1$.

\section{Line element}
\label{sec:lel}

It is very difficult to find physically relevant solutions of Einstein's equations. To simplify the resulting system of differential equations, one
usually assumes that certain conditions are satisfied which follow from the physical properties of the system under consideration. In the case of 
compact objects,  two assumptions are made, namely, stationarity and axial symmetry. The first condition means that the field does not depend explicitly
on the time coordinate, say $t$. This is in accordance with our experience since the shape and rotation of isolated compact objects have not been observed to change 
over long periods of time. The second condition is also based on observations. Indeed, all compact objects observed in Nature are characterized by a particular 
rotation with respect to an axis located inside the object. The rotation axis determines a privileged direction with respect to which compact objects 
are usually symmetric. It is therefore physically meaningful to assume that the gravitational field and, consequently, the spacetime metric are stationary and 
axially symmetric. Under these conditions, one can show that the line element  reduces to the Weyl-Papapetrou-Lewis form \cite{solutions}
\be
ds^2 = f(dt-\omega d\varphi)^2 - 
f^{-1}\left[e^{2 \gamma  }(d\rho^2+dz^2) +\mu^2 d \varphi^2\right] \ ,
\label{lel}
\ee
where $f$, $\omega$, $\mu$, and $\gamma$ are functions which depend on the spatial coordinates $\rho$ and $z$ only. Notice that the metric does not depend
on the coordinates $t$ and $\varphi$ as a consequence of the stationarity and axial symmetry assumptions, respectively. 

The corresponding field equations are, in general,  a highly non-linear and complicated system of second-order partial  
differential equations. Several representations for the field equations are known in the literature \cite{solutions,quev11}. 
The particular form of the line 
element (\ref{lel}) is convenient for the analysis of the field equations structure.  
  For instance, in the limiting case of static vacuum field $(\omega=0)$, the function
$f$ turns out to satisfy a second-order linear differential equation, whereas all the non-linearities of Einstein's equations are contained in the metric function 
$\gamma$. In the case of electrovacuum spacetimes, the function $\mu$ turns out to satisfy a harmonic differential equation so that it can be chosen as $\mu=\rho$ to simplify the form of the remaining field equations. In the general case of interior perfect-fluid metrics with electromagnetic field, the functions $f$, $\omega$ and 
$\mu$ satisfy a system of three coupled, non-linear, second-order, partial differential equations and the function $\gamma$ is determined by a set of two first-order 
differential equations which can be integrated by quadratures.

As we will see below, for the representation of particular exact solutions, however, the cylindrical-like coordinates of the line element (\ref{lel}) are not very convenient. In fact, 
the physical significance of particular solutions can be understood more easily  by using spherical-like coordinates. Also, the motion of test particles in the gravitational field of compact objects have been analyzed mostly in spherical-like coordinates \cite{pqr13}. 

An additional condition must be imposed in order for a particular solution to describe the exterior field of compact objects, namely, asymptotic flatness. This is a physical condition, implying that far away from the gravitational and electromagnetic source the fields must vanish, which is equivalent to saying that the metric reduces to the flat Minkowski metric. All the solutions presented in this work satisfy this physical condition.

\section{Exterior electrovacuum solutions}
\label{sec:ext}

The first vacuum solution of Einstein's equation was obtained by Schwarzschild in 1916. In spherical-like coordinates, it can be expressed as
\be
ds^2 = \left(1-\frac{2m}{r}\right) dt^2 - \frac{dr^2}{ 1-\frac{2m}{r} } - r^2(d\theta^2 + \sin^2\theta d\varphi^2) \ .
\ee
It represents the exterior field of a static spherically symmetric mass distribution. In this case, only one multipole moment is different from zero, namely, the monopole $M_0= m$. This is due to the spherical symmetry and the lack of rotation and electromagnetic field. The complexity of the field equations for the general stationary case is so high that it took almost fifty years to take into account the rotation of the source in the Kerr metric \cite{kerr63} 
\be
 ds^2=\frac{(\Delta -a^2\sin^2\theta)}{\Sigma
}dt^2-\frac{2a\sin^2\theta (r^2+a^2-\Delta)}{\Sigma}dtd\varphi \nonumber
\ee
\be
\label{kerr}
- \left[\frac{(r^2+a^2)^2-\Delta
a^2\sin^2\theta}{\Sigma}\right]\sin^2\theta
d\varphi^2-\frac{\Sigma}{\Delta}dr^2-\Sigma d\theta^2, 
\ee
where 
\be
\Delta = r^2-2m r+a^2 \ ,\quad \Sigma =r^2+a^2\cos^2\theta\ .
\ee
The parameter $a$ stands for the angular momentum per unit mass, $J/m$, as measured
by a distant observer. This can also be seen at the level of the multipole moments which in this case can be expressed as\footnote{Notice that the original Geroch-Hansen definition of multipole moments leads to expressions with the opposite sign in front of $M_n$ so that, for instance, the total mass is negative, $M^{GH}_0=-m$.
However, a conventional normalization of the multipoles can be performed so that a positive sign for the total mass is obtained. We use here this convention.} 
\be
M_{2k} = (-1)^{k}m a^{2k} \ , \quad M_{2k+1} = 0 \ , \quad J_{2k}=0\ ,\quad J_{2k+1} = (-1)^k m a^{2k+1} \ .
\ee
Notice that the odd mass multipoles and the even angular-momentum multipoles vanish identically as a result of the additional symmetry of the Kerr solution with respect to the equatorial plane $\theta=\pi/2$.

The charged generalization of the Kerr metric is obtained by considering the electromagnetic vector potential as the 1-form  
\be
A =-\frac{Q r}{\Sigma}\left( dt  - a\sin^2\theta  d\varphi \right) ,
\ee
which depends on the charge $Q$ and the specific angular momentum $a$. It then follows that the magnetic field is generated by the rotation of the charge 
distribution. To find the corresponding metric, it is necessary to solve the complete set of Einstein-Maxwell equations (\ref{emax1}) simultaneously. The final
expression for the solution turns out to be identical to (\ref{kerr}) and the only difference is at the level of the function $\Delta$, namely,
\be
\Delta = r^2-2m r+a^2 +Q ^2 \ .
\ee
The corresponding electromagnetic multipoles $E_n$ and $H_n$ turn out to proportional to $M_n$ and $J_n$, respectively, as a result of the non-linear gravitational 
interaction between the mass and the charge distribution.

According to the black hole uniqueness theorems \cite{heus96}, the charged Kerr solution is the most general electrovacuum solution that describes the gravitational and electromagnetic field of a black hole. This means that for a black hole all multipole moments must be given in terms of the mass ($M_0=m)$, the angular momentum ($J_1 = J)$, and the  electric charge ($E_0 = Q)$. All the remaining higher multipoles must depend explicitly on only these three parameters. In the case of other 
astrophysical compact objects, like white dwarfs or neutron stars, the uniqueness theorems are not valid. This means that higher multipoles, like the mass quadrupole or the electric dipole, could play an important role in the description of the gravitational field. 

To begin with, let us consider a static mass with a  quadrupole  
deformation. There are several solutions in the literature that can be used to describe a mass with a quadrupole  \cite{weyl17,erro59,ts73,gm85,quev87,quev11a}. The common
feature of most solutions is that they are given in terms of quite complicated analytical expressions. To our knowledge, the simplest solution for a mass with
a quadrupole is the $q-$metric \cite{quev11a}
\bea
&& ds^2  =  \left(1-\frac{2m}{r}\right) ^{1+q} dt^2  
- \left(1-\frac{2m}{r}\right)  ^{-q} \nonumber \\
&& \times \left[ \left(1+\frac{m^2\sin^2\theta}{r^2-2mr }\right)^{-q(2+q)} \left(\frac{dr^2}{1-\frac{2m}{r} }+ r^2d\theta^2\right) + r^2 \sin^2\theta d\varphi^2\right] ,
\label{zv}
\eea
which was originally found by Zipoy \cite{zip66} and Voorhees \cite{voor70} in prolate spheroidal coordinates. If the quadrupole parameter vanishes $(q=0)$, we
recover the spherically symmetric Schwarzschild solution. 

Static solutions with higher multipoles are also known. For instance, the most general Weyl solution in cylindrical-like coordinates (\ref{lel}) takes the form
\be
\ln f = 2 \sum_{n=0}^\infty \frac{a_n}{(\rho^2+z^2)^\frac{n+1}{2}} P_n({\cos\theta}) \ ,
\qquad \cos\theta = \frac{z}{\sqrt{\rho^2+z^2}} \ ,
\label{weylsol}
\ee
where $a_n$ $(n=0,1,...)$ are arbitrary constants, and $P_n(\cos\theta)$ represents the Legendre
polynomials of degree $n$. 
The expression for the metric function $\gamma$ 
can be calculated by quadratures. Then, 
\be
\gamma = - \sum_{n,m=0}^\infty \frac{ a_na_m (n+1)(m+1)}{(n+m+2)(\rho^2+z^2)^\frac{n+m+2}{2} }
\left(P_nP_m - P_{n+1}P_{m+1} \right) \ .
\ee
Although, in principle, the Weyl solution should contain all asymptotically flat static metrics, it is not very convenient for the investigation of the physical significance of the metrics. For instance, the Schwarzschild solution can be obtained by selecting the values of the parameters $a_n=a_n^S$ in such a way that the infinite sum 
\be
\sum_{n=0}^\infty \frac{a_n^S}{(\rho^2+z^2)^\frac{n+1}{2}} P_n({\cos\theta}) = \frac{1}{2}\ln\left(1-\frac{2m}{r}\right) \ .
\ee
converges to the Schwarzschild value. An alternative representation which is more suitable for the physical analysis of the metrics is given in terms of prolate spheroidal coordinates
\be
 ds^2 = f dt^2 - \frac{m^2}{f}\left[ e^{2\gamma}(x^2-y^2)\left( \frac{dx^2}{x^2-1} + \frac{dy^2}{1-y^2} \right) 
+ (x^2-1)(1-y^2) d\varphi^2\right] ,
\label{lelxy}
\ee
where $m$ is a constant. The general asymptotically flat vacuum solution can be written as \cite{erro59,quev89}
\be
 \ln f= 2 \sum_{n=0}^\infty (-1)^{n+1} q_n P_n(y) Q_n (x) \ , \quad q_n = const 
\label{gensolxy}
\ee
where $P_n(y)$ are the Legendre polynomials, and $Q_n(x)$ are the Legendre functions of second kind. The metric function $\gamma$ is quite cumbersome and
cannot be written in a compact form. The physical significance of the parameters $q_n$ can be obtained by calculating the corresponding Geroch multipole moments. In this case, we have that
\be
M_n = N_n + R_ n\ , \quad N_n = (-1)^n \frac{n! m^{n+1}}{(2n+1)!!} q_n \ ,
\ee
where $N_n$ represent the Newtonian multipole moments and $R_n$ are relativistic corrections which must be calculated explicitly for each value of $n$. For instance, 
$R_0=R_1=R_2=0$, $R_3 = - \frac{2}{5}m^2 N_1$, $R_4 = - \frac{2}{7} m^2 N_2 - \frac{6}{7} m N_1^2$, etc. We conclude that the parameters $q_n$ are the Newtonian multipoles, modulo a constant multiplicative factor, and determine also the relativistic corrections.

The Schwarzschild metric is a particular case of the general solution (\ref{gensolxy}) with $q_0=1$ and $q_k=0$ ($k>0$), and can be written as 
\be
f= \frac{x-1}{x+1}  \ , \quad \gamma = \frac{1}{2}\ln\frac{x^2-1}{x^2-y^2} \ .
\ee
It reduces to the standard form in spherical-like coordinates after applying the coordinate transformation $x=r/m-1$ and $y=\cos\theta$. Moreover, the $q-$metric 
corresponds to 
\be
f= \left(\frac{x-1}{x+1}\right)^{1+q}   \ , \quad \gamma = \frac{1}{2}(1+q)^2 \ln\frac{x^2-1}{x^2-y^2} \ ,
\ee
for which the leading mass multipoles are 
\be 
M_0= (1+q)m\ , \quad M_2 = -\frac{m^3}{3}q(1+q)(2+q)\ .
\ee
This shows that the parameter $q$ determines the quadrupole, but it also affects the total mass of the object. 

As for the stationary generalizations of the above static solutions, many of them have been obtained by using different solution generating techniques. 
For instance, the first generalization of the Kerr metric with an arbitrary mass quadrupole moment was obtained in \cite{qm85}.  Other stationary 
generalizations with quadrupole and higher moments were obtained in \cite{mn92,cm90,qm91,mms00,prs06}. The general form of all these metrics does not 
allow to express them in a simple manner. All the metric that generalizes the stationary Kerr metric are expected to be equivalent at the level of the 
quadrupole moment, up to a redefinition of the parameters that determine the mass quadrupole.

This short review of electrovacuum solutions shows that the situation is not complicated when  we limit ourselves to the case of a rotating charged mass.
This is due to the uniqueness theorems according to which the charged Kerr metric can be used to described any black hole in general relativity. 
Once we pretend to take into account the effects of a mass quadrupole, the situation becomes more and more complicated due to the increasing number
of exact solutions. As mentioned above, one expects that all these solutions are equivalent at the level of the quadrupole moment due to the 
deformation and the rotation of the source. We believe that other physical conditions should be imposed in order to establish the difference between
all available exact solutions. Further work in this direction is necessary in order to determine the physical relevance of all the solutions available in
the literature.


\section{Interior solutions}
\label{sec:int}

In the previous section, it was shown that in principle it is possible to describe the exterior field of compact objects by using exact electrovacuum solutions
of Einstein-Maxwell equations. Each exterior solution, however, should be matched at the surface of the compact object with a physically meaningful interior solution.
Consider, for instance, the simplest exterior solution with a mass monopole, i.e., the exterior Schwarzschild solution. If we suppose a perfect-fluid model for 
the interior counterpart with constant energy density ($\rho=\rho_0=const$) and no charge distribution $(F_{\mu\nu}=0)$, the resulting field equations can be integrated 
analytically, leading to the interior metric
\be
ds^2 = \left[\frac{3}{2}f(R) - \frac{1}{2}f(r)\right]^2 dt^2 - \frac{dr^2}{f^2(r)} - r^2(d\theta^2 + \sin^2\theta d\varphi^2)\ ,
\ee
with 
\be
f(r)= \sqrt{1-\frac{2mr^2}{R^3}}\ .
\ee
Moreover, the pressure of the perfect fluid turns out to be a function of the radial coordinate $r$ only 
\be
p = \rho_0 \frac{f(r) - f(R)}{3f(R)-f(r)}\ .
\ee
This solution satisfies the matching conditions at the surface of a sphere of radius $R$. From this point of view, it is a good candidate for describing the interior of a spherically symmetric mass distribution. From a physical point of view, however, this solution is not acceptable. Indeed, the assumption of constant inner density implies that the fluid is incompressible which leads to an infinite speed of sound. There are many other static perfect-fluid solutions (see \cite{solutions} for a list of the most relevant solutions). Nevertheless, none of them has a  meaningful physical significance. The generalizations that include  charge distributions present similar problems. 

In the case of rotating fields, the situation is similar. Einstein's field equations have no known and physically acceptable
interior solution that could be matched to the exterior Kerr metric. In
particular, there are no interior solutions that could represent objects like
the Earth or other rigidly rotating astrophysical objects. This is a major problem in general relativity.

There exist some exact interior solutions of Einstein's equations with a perfect-fluid source equipped with an electromagnetic field \cite{ggq13} that satisfy 
all the energy conditions and are well-behaved in the entire spacetime. They are interpreted as describing the gravitational and electromagnetic fields of disk-halos. 
However, this kind of solutions cannot be matched with any known exterior solution.

One century after the discovery of the exterior Schwarzschild solution, we see that even the simplest case of an interior field with only mass monopole moment does not have a definite solution in general relativity. In view of this situation, it seems convenient to try new and different approaches. We propose the following idea. 
The charged Kerr metric describes the exterior field of black holes. Once a particle crosses the horizon of a black hole, according to classical general relativity, 
it undoubtedly must end at the singularity. This means that the interior of black hole is a singularity where the classical theory breaks down. Consequently, it is not possible to describe the interior field of a black hole by using only classical general relativity; instead, it should be a problem of quantum gravity. In fact, we know that a crucial test of any  quantum gravity model must be the avoidance of the classical singularities. Consequently, the interior counterpart of the charged Kerr metric cannot be found by using Einstein's equations only. 

Consider now classical compact objects, not including black holes. In the static vacuum case, we usually assume that they are described by the spherically symmetric Schwarzschild metric, the same which is used for black holes. This assumption should be changed. Indeed, it is hard to imagine in Nature a completely spherically symmetric compact object. Therefore, it is necessary to take into account the natural deviations from spherical symmetry. The simplest way to reach this end is to 
consider the contribution of an axially symmetric quadrupole moment.  As we mentioned in the previous section, there are several exterior metrics which represent the
gravitational field of mass with quadrupole. The search for the corresponding interior counterparts with quadrupole moment will certainly contribute to understand the difference between the exterior metrics. It could be, for instance, that each exterior metric corresponds to an interior metric with a particular physical structure.
Also, if it turns out that  an explicit exterior metric does not allow the existence of a reasonable interior counterpart, it should not be considered as physically relevant. 

We already started a program in which the main goal is to find interior solutions with quadrupole. The starting point is the exterior $q-$metric presented in the previous section. To search for the corresponding interior metric, it is very important to choose a convenient line element because the structure of the field equations depends on its explicit form. 
In \cite{qt15}, we proposed the following line element 
\be
ds^2 = e^{2\psi} dt^2 - e^{-2\psi}\bigg[e^{2\gamma}\left(\frac{dr^2}{h} + d\theta^2\right) + \mu^2 d\varphi^2\bigg]\ ,
\label{lel1}
\ee
where $\psi=\psi(r,\theta)$, $\gamma=\gamma(r,\theta)$, $\mu=\mu(r,\theta)$, and $h=h(r)$. For a perfect-fluid source, the main field equations can be written as 
\be
\frac{\mu_{,rr}}{\mu} +  \frac{\mu_{,\theta\theta}}{h\mu} +\frac{h_{,r}\mu_{,r}}{2h\mu} = \frac{16\pi}{h} p  e^{2\gamma - 2\psi}\ ,
\label{mu}
\ee 
\be
\psi_{,rr} + \frac{\psi_{,\theta\theta}}{h} +\left(\frac{h_{,r}}{2h}+\frac{\mu_{,r}}{\mu}\right) \psi_{,r} + \frac{\mu_{,\theta} \psi_{,\theta}}{h\mu} 
= \frac{4\pi}{h}(3  p + \rho)  e^{2\gamma - 2\psi} \ .
\label{psi}
\ee
Moreover, the metric function $\gamma$ is determined by two first-order differential equations that can be integrated once the remaining metric functions are known
explicitly. This is an advantage of the above line element. An additional advantage is that the conservation of the energy-momentum tensor leads to two simple expressions
\be 
p_{,r} = -(\rho+p)\psi_{,r}  \ ,\quad p_{,\theta} = -(\rho+p)\psi_{,\theta}\ ,
\label{claw}
\ee
which resemble the Tolman-Openheimer-Volkov relation of the spherically symmetric case \cite{solutions}. This is particularly important when trying to perform the integration of the main field equations. It is still very difficult to solve the above system of differential equations. We therefore propose to use solution generation techniques. The first step in this direction was taken in \cite{qt15} where a transformation was derived by means of which it is possible to generate interior 
solutions with quadrupole moment, starting from spherically symmetric interior solutions. The investigation of the resulting axially symmetric solutions with quadrupole is currently under investigation.


\section{Conclusions}
\label{sec:con}

In this work, we presented a brief review of the problem of describing the gravitational and electromagnetic field of astrophysical compact objects by using
the multipole structure of exact solutions of the Einstein-Maxwell equations. In the case of the exterior field, we presented  the properties of the main solutions and its multipole moments.
Using the black hole uniqueness theorems, we observe that the charged Kerr solution is the only metric which contains the mass monopole, the dipole angular momentum and the
charge monopole. Once higher multipoles are taken into account, the uniqueness theorems are no more valid and several solutions are available in the literature. We mention the $q-$metric as the simplest solution with quadrupole moment. 
We showed that there exist general static solutions with an infinite number of parameters which determine the Newtonian and the relativistic Geroch-Hansen multipole moments. We mentioned that it is possible to find the corresponding stationary electrovacuum generalizations by using solution generating techniques. In this manner, one can say that the problem of describing the exterior field of astrophysical compact objects can be solved by using multipole moments.

In the case of interior solutions, the situation is completely different. Even the simplest case of perfect-fluid source with only mass monopole cannot be solved 
completely. A major problem, for instance, is that there is no known physically meaningful interior solution that could be matched with the exterior Kerr metric.
We therefore propose an alternative approach. The interior counterpart of the charged Kerr solution cannot be found in classical general relativity because inside the
horizon a curvature singularity exists which implies the break down of the classical theory. Quantum gravity should be used to investigate the internal structure of 
black holes. As for other compact objects, it is necessary to take into account the natural deviations from spherical symmetry by adding higher multipoles. In particular, we propose to use the exterior $q-$metric to search for an interior solution with quadrupole. Preliminary calculations show that it is possible to 
find interior solutions by using a particular transformation which allows one to generate interior axially symmetric solutions with quadrupole moment. This task is currently under investigation.

\subsection*{Acknowledgements}
This work was  supported by DGAPA-UNAM, Grant No. 113514, Conacyt-Mexico, Grant No. 166391, and 
MES-Kazakhstan, Grant No. 3098/GF4.


\begin{thebibliography}{99}

\bibitem{eins15} A. Einstein, {\it Die Feldgleichungen der Gravitation.} Sitzungsberichte der Preussischen Akademie
der Wissenschaften zu Berlin, 844-847 (1915).

\bibitem{solutions}
H. Stephani, D. Kramer, M. A. H. MacCallum, C. Hoenselaers, and E. Herlt,
{\it Exact Solutions of Einstein's Field Equations} (Cambridge University Press, Cambridge, UK, 2003).

\bibitem{quev11} H. Quevedo, {\it Multipolar solutions}, 
in Cosmology and Gravitation, Proceedings of the XIVth Brazilian School of Cosmology and Gravitation, edited M. Novello and S. E. Perez Bergliaffa 
(Cambridge Scientific Publishers, 2011) pp. 83 – 97.

\bibitem{pqr13} D. Pugliese, H. Quevedo and R. Ruffini, {\it Equatorial circular orbits of neutral test particles in the Kerr-Newman spacetime.},
Phys. Rev. D {\bf 88}, 024042 (2013).


\bibitem{ger70} R. Geroch, {\it Multipole moments. I. Flat space.}
 J.\ Math.\ Phys. {\bf 11}, 1955 (1970);
{\it Multipole moments. II. Curved space.}
 J.\ Math.\ Phys. {\bf 11}, 2580 (1970).

\bibitem{hans74} R. O. Hansen, 
{\it Multipole moments of stationary spacetimes.}
J.\ Math.\ Phys. {\bf 15}, 46 (1974).

\bibitem{quev90} H. Quevedo, {\it
Multipole Moments in General Relativity -Static and Stationary Solutions-}, Fort. Phys. {\bf 38}, 733 - 840 (1990).

\bibitem{quev92} H. Quevedo, {\it
Generating Solutions of the Einstein - Maxwell Equations with Prescribed Physical Properties.}, Phys. Rev. D {\bf 45}, 1174 - 1177 (1992).

\bibitem{kerr63} R. P. Kerr, {\it  Gravitational field of a spinning mass as an example of algebraically
special metrics} Phys. Rev. Lett. {\bf 11}, 237 (1963).

\bibitem{heus96} M. Heusler, {\it Black Hole Uniqueness Theorems} (Cambridge University Press, Cambridge, UK, 1996). 

\bibitem{weyl17} H. Weyl, {\it Zur Gravitationstheorie.} Ann. Physik (Germany) {\bf 54}, 117 (1917).

\bibitem{erro59} G. Erez and N. Rosen, {\it The gravitational field of a particle possessing a quadrupole moment.}
Bull. Res. Counc. Israel {\bf 8}, 47 (1959).

\bibitem{ts73} A. Tomimatsu and H. Sato, {\it New series of exact solutions for gravitational
fields of spinning masses} Prog. Theor. Phys. {\bf 50}, 95 (1973). 

\bibitem{gm85} T. I. Gutsunaev  and V. S. Manko, {\it On the gravitational field of a mass possessing
a multipole moment.} Gen. Rel. Grav. {\bf 17}, 1025 (1985). 

\bibitem{quev87} H. Quevedo, {\it On the exterior gravitational field of a mass with a multipole moment.} Gen. Rel. Grav. {\bf 19}, 1013 - 1023 (1987).

\bibitem{quev11a} H. Quevedo, {\it Exterior and interior metrics with quadrupole moment.} Gen. Rel. Grav. {\bf 43}, 1141-1152 (2011).


\bibitem{zip66} D. M. Zipoy, 
{\it Topology of some spheroidal metrics.}
 J. Math. Phys. {\bf 7}, 1137 (1966).

\bibitem{voor70} B. Voorhees, 
{\it Static axially symmetric gravitational fields.}
 Phys. Rev. D {\bf 2}, 2119 (1970).

\bibitem{quev89} H. Quevedo, {\it General static axisymmetric solution of Einstein's vacuum field equations in prolate spheroidal coordinates.} Phys. Rev. D
{\bf 39}, 2904 - 2911 (1989).

\bibitem{qm85} H. Quevedo and B. Mashhoon, {\it Exterior gravitational field of a rotating deformed mass.} Phys. Lett. A {\bf 109}, 13 - 18 (1985).

\bibitem{mn92} V. S.Manko and I.D. Novikov, 
Class. Quantum Grav. {\bf 9}, 2477 (1992).

\bibitem{cm90} J. Castejon-Amenedo and V. S. Manko, {\it On a stationary rotating mass with an arbitrary multipole structure.} Class. Quantum Grav. {\bf 7},  
779-785 (1990). 

\bibitem{qm91} H. Quevedo and B. Mashhoon, {\it Generalization of Kerr spacetime.} Phys. Rev. D {\bf 43}, 3902-3906 (1991).

\bibitem{mms00} V. S. Manko, E. Mielke, and J. D. Sanabria, {\it Exact solution for the exterior field of a rotating neutron star.} Phys. Rev. D {\bf 61},
081501 (2000). 

\bibitem{prs06} L. Pach\'on, J. A. Rueda, J. D. Sanabria, {\it Realistic exact solution for the exterior field of a rotating neutron star.}
Phys. Rev. D {\bf 73}, 104038 (2006).

\bibitem{ggq13}  A. Guti\'errez-Pi\~neres, G. Gonz\'alez and H. Quevedo, {\it 
Conformastatic disk-haloes in Einstein-Maxwell gravity.} Phys. Rev. D {\bf 87}, 044010 (2013).

\bibitem{qt15} H. Quevedo and S. Toktarbay, {\it Generating static perfect-fluid solutions of Einstein's equations.}
J. Math. Phys. {\bf 56}, 052502 (2015).

\end{thebibliography}
\end{document}